\documentclass[prb,twocolumn]{revtex4}
\usepackage{graphicx}
\usepackage{epsfig}
\usepackage{amssymb,latexsym,amsmath}

\usepackage{hyperref}

\usepackage{ulem}
\usepackage{color}
\usepackage{framed}

\begin{document}

\title{ \Large{ \bf Comparative study of heat-driven and power-driven refrigerators with Coulomb-coupled quantum dots }}

\author{A.-M. Dar\'{e}}\email{Anne-Marie.Dare@univ-amu.fr} 
\affiliation{ Aix-Marseille Universit\'e, CNRS, IM2NP UMR 7334, \small \it  13397, Marseille, France}

\date{\today}
\begin{abstract}

Multiterminal multidot devices have been put forward as versatile and high-performing setups for
thermoelectric energy harvesting at the nanoscale.
With a technique that encompasses and overtakes several of the usual theoretical tools used in this context, we 
explore a three-terminal Coulomb-coupled-dot device for refrigeration purposes.
The refrigerator is monitored by either a voltage or a thermal bias.
This comparative study shows that the heat-driven refrigerator is underperforming relative to
the power-driven one, due to scarce on-dot charge fluctuations.

\end{abstract}

\maketitle

\section{Introduction}

Thermoelectrics is a promising candidate for energy harvesting development.
The investigation of thermoelectric properties at the nanoscale has taken a leap forward recently. 
It was sparked partly by a famous paper by Mahan and Sofo [\onlinecite{Mahan96}], demonstrating that confinement and energy filtering that are features 
of nanoscale systems can boost the thermoelectric figure of merit.
As an example Coulomb blockade dots coupled by tunneling, or capacitively, can be nearly optimal energy converters both in the two-terminal and three-terminal environments [\onlinecite{Jordan13, Sothmann15,Josefsson18}]. 
Nanoscale thermal machines for refrigeration with quantum dots (QD) experience also a significant
development [\onlinecite{Edwards93, Prance09, Muhonen12, Pekola14, Feshchenko14}].

\begin{figure}[h!]
\begin{center}
\includegraphics[width=.48\textwidth]{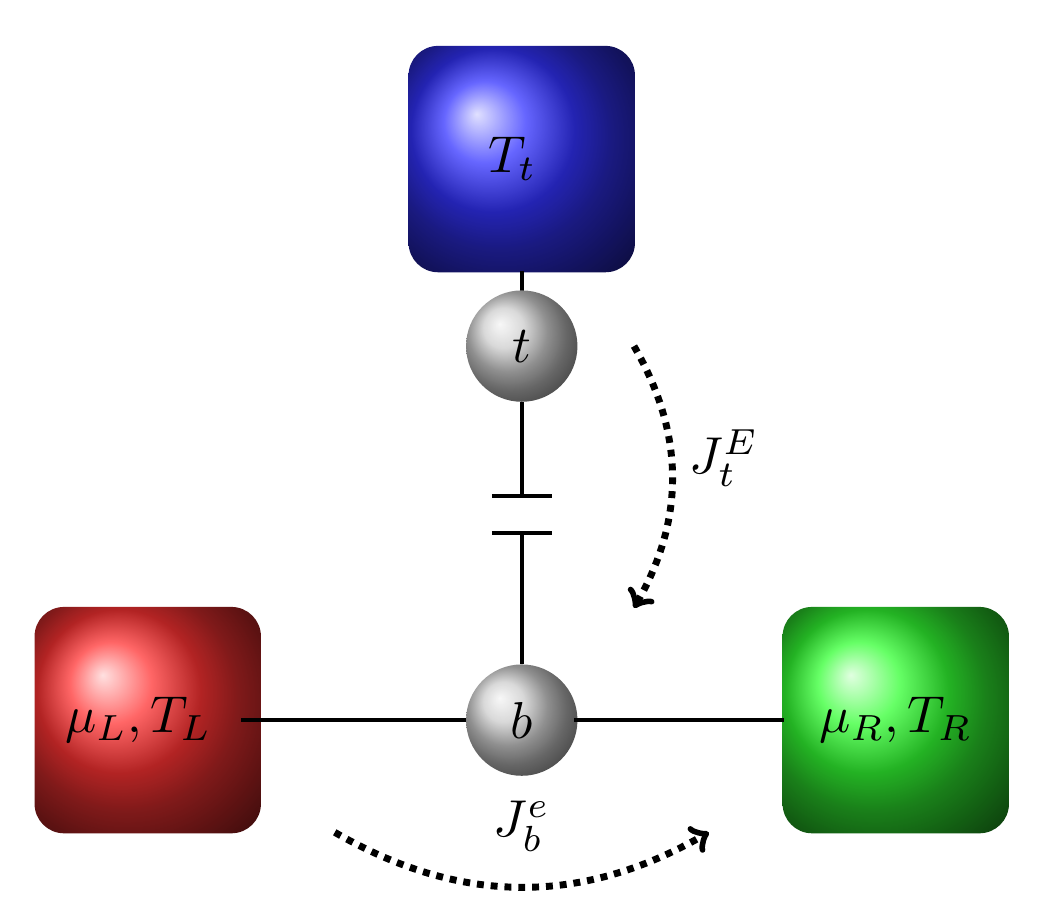}
\caption{Schematic representation of the device: the $t$ dot (top) is connected to a cold reservoir to be cooled, and coupled by Coulomb interaction to a 
 $b$ dot (bottom). The latter is connected to two reservoirs (left and right) through which voltage and thermal biases can be applied.}
 \label{figdispo}
\end{center}
\end{figure}

In the present paper we study a mesoscopic system consisting of two quantum dots and three electronic reservoirs as illustrated in Fig.~\ref{figdispo}. The dots are 
capacitively coupled by Coulomb repulsion.
This device was conceived by S\'anchez and B\"uttiker in Ref.~[\onlinecite{Sanchez11}]. It is quite versatile, and has been suggested to realize 
an engine [\onlinecite{Sanchez11, Thierschmann16, Zhang16,Dare17}], for refrigeration [\onlinecite{Zhang15, Benenti17, Erdman18}], 
for thermal control of charge current [\onlinecite{Thierschmann16, Thierschmann16-2}], and for thermal diode and transistor engineering [\onlinecite{Thierschmann16,Sanchez17,Sanchez17-2}].
Recently this setup was also proposed as a nanoscale thermometer [\onlinecite{Zhang19}].
One of its main appeal is the actual decoupling of charge and heat currents, which constitutes a promising way to high-performing
devices [\onlinecite{Mazza15, Thierschmann16-2}].

The experimental side is not to be outdone, and the first realization of the two-dot three-terminal device in the nanoengine regime is due to
Thierschmann {\it et al.}, a work published in Ref.~[\onlinecite{Thierschmann15-1}] and reviewed in Ref.~[\onlinecite{Thierschmann16-2}]. It was also 
experimentally investigated for thermal gating [\onlinecite{Thierschmann15-2}].
Additionally, a very similar device was recently conceived as the first experimental autonomous Maxwell demon [\onlinecite{Koski15, Koski18}] and further 
studied theoretically [\onlinecite{Kutvonen16}].
More broadly, in the buoyant field of nanoscale thermoelectrics, other kinds of nano-devices have been recently examined, essentially for energy harvesting purpose, heat diode realization, or in the Maxwell 
demon context, both 
experimentally [\onlinecite{Roche15, Jaliel19,Thierschmann13}] and theoretically, [\onlinecite{Ruokola11, Zhang17, Zhang18, Bhandari18, Jiang18, Yang19, Sanchez19,Jing18,Mazal19, Daroca18}]. 
A short pedagogical review can be found in Ref.~[\onlinecite{Whitney19}].
The device sketched in Fig.~\ref{figdispo} will be studied here for refrigeration purpose, and two different settings will be analyzed and compared.
First one: by applying a thermal bias between the two bottom reservoirs, an all-thermal refrigerator, without any electric power ($\mu_L = \mu_R$), can be realized.
This kind of all-thermal machine is also sometimes called autonomous [\onlinecite{ambiguite}], absorption [\onlinecite{Correa14,Segal18,Mitchison19,Mitchison19-2}], self-contained 
[\onlinecite{Linden10}], or cooling by heating refrigerator [\onlinecite{Benenti17}] (and references therein).
All-thermal refrigerator has a long history in thermodynamics, dating back to 1857 where it was invented by Carr\'e. However, its first 
quantum experimental 
release, with three trapped ions, is very recent [\onlinecite{Maslennikov19}].
The initial paper mentioning all-thermal refrigeration within the present two-dot three-terminal setup is to our knowledge  by Benenti {\it et al.} 
[\onlinecite{Benenti17}]. This suggestion was soon implemented by Erdman {\it et al.} [\onlinecite{Erdman18}].
Other all-thermal quantum refrigerator devices have been explored, for example, devices involving a small number of qubits or qutrits [\onlinecite{Linden10,Mukhopadhyay18}], devices made of four 
QD [\onlinecite{Venturelli13}] or implying three levels coupled to bosonic baths [\onlinecite{Correa14, Segal18}]. See Ref.~[\onlinecite{Mitchison19, Mitchison19-2}] and
references therein for other implementations. 
Second setting: we shall consider the same device devised as an electric refrigerator, namely monitored by a voltage bias $eV= \mu_L-\mu_R$, applied between the two bottom 
leads [\onlinecite{Zhang15}]. 

The heat- and power-driven refrigerator properties have already been investigated in Refs.~[\onlinecite{Zhang15, Erdman18}], though in a $T$-matrix quantum master 
equation 
limited to sequential tunneling processes (SQME). While this approximation is believed to be valid for weak dot-lead tunnel couplings,
higher orders as cotunneling events can become quantitatively important [\onlinecite{Bhandari18}] even for weak coupling, particularly 
as shown recently close to the maximum efficiency regime [\onlinecite{Josefsson19}].
Similarly even in the weak-coupling situation, broadening as well as energy shifts can have a quantitative impact on performances.
Furthermore although strong tunnel coupling would be detrimental to filtering and thus efficiency, it can be beneficial to power, and is sometimes 
realized in experimental setups:
for example in Ref. [\onlinecite{Thierschmann15-1}], temperatures are of the same order of magnitude as tunnel couplings.
In the same kind of device, yet in the context of Coulomb drag without thermal bias, the regime where tunneling coupling is much higher that 
temperature has been considered [\onlinecite{Keller16}].
Besides, for nano-device cooling purpose, exploring the low-temperature regime where the weak-coupling assumption can be ruled out, is a topic of interest [\onlinecite{Benenti17}].
These issues demonstrate the usefulness of developing a framework to access the strong coupling regime
beyond SQME, and even beyond QME that includes cotunneling. There are not so many methods to address these operating regimes.
One can cite a numerical approach used by Bhandari {\it et al.} in Ref. [\onlinecite{Bhandari18}], a Keldysh non-equilibrium Green's function method
including the one bubble correction beyond Hartree in self-energy, recently developed in a closely related four terminal device [\onlinecite{Sierra19}], 
and a non-crossing approximation (NCA), which has been applied to the present device for the engine appliance [\onlinecite{Dare17}]. 

We use the NCA in the current work for refrigeration purpose, we will show that the performances of the two types of refrigerator are very different, due to on-dot
charge fluctuations that are rather scarce for the all-thermal setup. If the latter is not very efficient and cannot be realized at too low temperature, the electric 
refrigerator is rather high-performing.
The outline is as follows: after a presentation of our model and method in Sec. II, we address the case of the all-thermal refrigerator in Sec. III, before
the electric one in Sec. IV. Summary and conclusions are displayed in Sec. V.

\section{Model}

The dots are indexed by $t$ or $b$ for top or bottom, and due to a strong local Coulomb repulsion they are described by a single nondegenerate orbital each.
They are coupled together through a nonlocal Coulomb repulsion $U$. This interaction is schematically represented in Fig.~\ref{figdispo} by a capacitive coupling, not allowing any charge transfer.
The three reservoirs [respectively top ($t$), left bottom ($L$), and right bottom ($R$)] are supposed to be equilibrium noninteracting Fermi seas, 
with their own chemical potentials and temperatures.
The Hamiltonian describing the present device can be written as $H = H_0+H_T$, where the disconnected part for dots and leads reads, in usual notations
\begin{equation}
H_0=  \epsilon_t \hat{n}_t+ \epsilon_b \hat{n}_b+ U \hat{n}_t \hat{n}_b +\sum_ {\alpha=t,L,R}H_{0\alpha }  \ ,
\label{h0}
\end{equation}
with $H_{0\alpha }= \sum_{k} \epsilon_{k \alpha}c^\dagger_{k \alpha} c_{k \alpha}$, 
$\hat{n}_b = d_b^\dagger d_b$, and $\hat{n}_t = d_t^\dagger d_t$.
Hybridization between dots and leads reads
\begin{equation}
H_T =\sum_{k} \Bigl( V_{k t} c^\dagger_{k  t} d_t +hc \Bigr)
+\sum_{\beta=L, R}\sum_{k} \Bigl( V_{k \beta} c^\dagger_{k \beta} d_b +hc \Bigr)  \ .
\label{ht}
\end{equation}
We choose as frequently used, the hybridization parameters to depend only on the energy: $V_{k \alpha}=V_\alpha(\epsilon_{k \alpha})$ 
[\onlinecite{JauhoWingreenMeir1994}].
In the Keldysh Green's function formalism [\onlinecite{JauhoWingreenMeir1994}], the stationary charge  and energy currents flowing outside an $\alpha$ reservoir into a dot can be expressed as
\begin{eqnarray}
 \left( \begin{array}{c} J^e_\alpha \\ J^E_\alpha \end{array} \right)
& & = 
\frac{i }{\hbar} \int  \frac{d \epsilon}{2 \pi}
\left( \begin{array}{c} e \\ \epsilon \end{array} \right)
 \Gamma_\alpha(\epsilon)  \nonumber \\
& &\times \biggl[  f_\alpha (\epsilon)
G_d^>(\epsilon) + [1- f_\alpha (\epsilon)] G_d^<(\epsilon)   \biggr] ,
\label{alpha-currents}
\end{eqnarray}
where $G_d^\lessgtr(\epsilon)$ are the lesser and greater dot Green's functions, $f_\alpha (\epsilon) =(e^{(\epsilon-\mu_\alpha)/(k_B T_\alpha)}+1)^{-1}$ is the Fermi function of the $\alpha$ reservoir, and  
$\Gamma_\alpha(\epsilon)=2 \pi \rho_{\alpha}(\epsilon) |V_\alpha(\epsilon) |^2$ is the effective dot-lead hybridization function with 
$\rho_\alpha(\epsilon)$ the lead density of states, $e>0$ is the elementary charge.
In the integrands of Eq.~(\ref{alpha-currents}), the two terms can be interpreted as a balance between in and out currents flowing between the dot and the $\alpha$ lead, indeed, for fermions $i G_d^>(\epsilon) \ge 0$, whereas 
$i G_d^<(\epsilon) \le 0$.
In general in and out currents are much larger than the difference. 

The electric current flowing through the bottom part of the device will be expressed from its symmetric expression, with the convention of positive contribution of electrons traveling from left to right:
$J^e_b= J^e_{L}=-J^e_{R}=(J^e_{L}-J^e_{R})/2$, leading to
\begin{align}
J^e_b =&\frac{i e}{2\hbar} \int  \frac{d \epsilon}{2 \pi}  \biggl[   G_b^>(\epsilon)
\Bigl(\Gamma_{L}(\epsilon) f_{L}(\epsilon) - \Gamma_{R}(\epsilon) f_{R}(\epsilon)\Bigr) \nonumber \\
+
& G_b^<(\epsilon)
\Bigl( \Gamma_{L}(\epsilon) \bigl(1-f_{L}(\epsilon)\bigr) - \Gamma_{R}(\epsilon) \bigl(1-f_{R}(\epsilon)\bigr)\Bigr)  \biggr] ,
\label{courantelec}
\end{align}
where $G_b^\lessgtr(\epsilon)$ are the Green's functions for the bottom dot.
Finally the heat currents are defined by
\begin{equation}
J^Q_\alpha = J^E_\alpha -\frac {\mu_\alpha}{e} J^e_\alpha \ .
\end{equation}
In the present convention, heat currents are positive for heat extracted from the involved reservoir.
In the refrigerator device - it is also the case for the three-terminal two dot engine [\onlinecite{Sanchez11}] - the hybridization functions 
between the $b$ dot and the connected reservoirs, $\Gamma_L(\epsilon)$ and  $\Gamma_R(\epsilon)$, must be different and not proportional. 
As a consequence the usual simplification [\onlinecite{JauhoWingreenMeir1994}] that allows to calculate only the spectral function 
$A_b(\epsilon) = i[G_b^>(\epsilon)-G_b^<(\epsilon)]$, 
and from which emerges only the difference of Fermi functions, leading
 to a Landauer-like formula, will not apply for the electric current  in Eq. (\ref{courantelec}). 
It does not apply either to the heat current extracted from the $t$ reservoir, which reads
\begin{equation}
J^Q_t = \frac{i}{\hbar} \int \frac{d \epsilon}{2 \pi}  \epsilon \ \Gamma_t(\epsilon) \Bigl( f_t G_t^>(\epsilon) +\bigl( 1-f_t(\epsilon) \bigr) G_t^<(\epsilon) \Bigr) ,
\end{equation}
where $G_t^\lessgtr(\epsilon)$ are the Green's functions for the $t$ dot.
In the following we choose
\begin{align}
\Gamma_{t}(\epsilon) &=  \Gamma_t \nonumber \\
\Gamma_{R}(\epsilon) &=\Gamma_b\  \theta(\epsilon-\epsilon_\Gamma)  \nonumber\\
\Gamma_{L}(\epsilon) &= \Gamma_{R} (-\epsilon) ,
\label{nosGamma}
\end{align}
with $\theta(\epsilon-\epsilon_\Gamma)$ the Heaviside function starting at the boundary $\epsilon_\Gamma = \epsilon_b +\frac U 2$. 
Engineering this kind of tunneling functions may be realized by making use of an additional quantum dot, or using metallic island as proposed 
recently [\onlinecite{Erdman18}].
In the following, $\Gamma_b= \Gamma_t =\Gamma$ will be the energy unit. We take $k_B=1$ and $e=1$. Without limiting the generality of the foregoing, we choose $\mu_t=0$. 
In the present model, the number of parameters is already large: three temperatures, two dot levels $\epsilon_b$ and $\epsilon_t$, as well as the 
Coulomb repulsion $U$. Steering these parameters enables to browse different regimes (engine, thermal gating, refrigeration, etc.). 
Concerning experimental devices, these parameters can be tuned by applying gate voltages and by modifying the distance between the dots.

Let us emphasize that too simple treatment that neglects fluctuations, such as static mean-field approach, cannot address the properties of the 
present or related devices [\onlinecite{Dare17,Yadalam19}]. 
To our knowledge the three-terminal two-dot thermal machine was only studied, except in Ref. [\onlinecite{Dare17}], in the framework of QME, with [\onlinecite{Walldorf17}] or without [\onlinecite{Erdman18, Zhang15}] cotunneling corrections. 
To calculate the Green's functions to obtain the currents, we use a non-crossing approximation [\onlinecite{Bickers87}], 
which is a simple current-conserving approximation [\onlinecite{Wingreen94}], and which has led to useful insights in the context of the Anderson impurity model, notably 
predicting the Kondo resonance and its energy scale. It is a fictitious particle technique [\onlinecite{Haule01}] that was readily extended in the Keldysh formalism to study non-equilibrium 
properties [\onlinecite{Wingreen94,Hettler98}].
This approximation is valid for $U \gg \Gamma$, better for high orbital degeneracy, but there is no restriction concerning temperatures compared to hybridization $\Gamma$, except at 
temperature much lower than the Kondo temperature.
The NCA was initially designed to study the infinite $U$ situation, and later extended to consider finite Coulomb repulsion by including vertex 
corrections [\onlinecite{Pruschke89,Keiter90,RouraBas09,Otsuki06}]. For the problem at hand, we need and use a finite $U$ version of the NCA, 
but we do not take into account these vertex corrections. 
Indeed going beyond, by developing the one crossing approximation (OCA) would involve significant numerical effort in the present out-of-equilibrium 
regime as detailed in Ref. [\onlinecite{Dare17}]. Furthermore, OCA is not a universal panacea [\onlinecite{Ruegg13}].
The present way to apply the finite-$U$ NCA is not flawless as raised in Refs.~[\onlinecite{Sposetti16,Daroca18}], however, 
the explored parameter regimes are such that we keep away from the region where severe problems such as underestimation of the Kondo 
resonance temperature arise [\onlinecite{Kotliar06}]. 

In the present approach the four non-equilibrium Green's functions ($G^{\lessgtr}_{b, t}$), characterizing the bottom and top dots are expressed 
in terms of eight Green's functions for four pseudoparticles, which are coupled and calculated self-consistently. The details of the 
self-consistent expressions were reported in Appendix A of Ref.~[\onlinecite{Dare17}]. 
The NCA is able to capture the atomic limit when $\Gamma \rightarrow 0$, and as a consequence in this limit, it encompasses QME that includes cotunneling, as shown 
in Ref.~[\onlinecite{Dare17}] for the engine setup.
Electric and thermal currents that are tied by conservation demands, depend on tiny details of Green's functions. 
In addition self-consistent calculations of the latter give results that are not very intuitive. As a consequence the numerical results will be hardly substantiated by analytical behaviors. 
In the present formalism, we calculate only averages of heat and charge currents, however, current fluctuations, which have been analyzed in this kind of setup [\onlinecite{Yadalam19,Cuetara11, Sanchez12, Sanchez13}], manifest themselves as will be discussed for the electric refrigerator.

From the heat currents, we can also readily evaluate the entropy production rate in the three reservoirs. It reads
$\dot{S}_0 = -\frac{J^Q_t}{T_t}-\frac{J^Q_L}{T_L}-\frac{J^Q_R}{T_R} $. Using the first principle and the heat current definition, 
we can rewrite it in the present notations as
\begin{equation}
\dot{S}_0 = J^Q_L \bigl(\frac{1}{T_R}-\frac{1}{T_L}\Bigr) +J^Q_t \Bigl(  \frac{1}{T_R}-\frac{1}{T_t}\Bigr) +\frac {(\mu_L-\mu_R) J^e_b/e}{T_R} .
\label{dotS0}
\end{equation}
This expression will be specified in the following for the two types of refrigerator. 
The NCA satisfies energy and charge conservation. In our calculations, we have checked that the second law $\dot{S}_0 >0$ is also fulfilled. 
This is not straightforward: for example the second principle may be violated in some local master equation approach [\onlinecite{Levy14}].

\section{All-thermal refrigerator}

An all-thermal refrigerator can be realized without work injection: to transfer heat from a cold source to a hot one, 
heat supplied by a source even hotter than the previous two can substitute to the injected work.  
For the present three sources indexed by $L$, $R$ and $t$, in descending order of temperature, the thermal machine will be a refrigerator if  a positive $J^Q_L$ can 
trigger a positive $J^Q_t$, while in accordance with the first principle $J^Q_R =-J^Q_t-J^Q_L$ will be negative. The coefficient
of performance (COP) is defined by the ratio ${J^Q_t}/{J^Q_L}$ and is bounded from above by the one ascribed to a reversible process
\begin{equation}
COP \le \frac{T_t}{T_R-T_t} \Bigl(1 -\frac{T_R}{T_L}  \Bigr) .
\end{equation}
The reversible COP is the product of the efficiency of an engine whose heat sources are the hot and warm reservoirs, times the COP of a refrigerator
 operating between the warm and cold reservoirs. 
The COP bound of an all-thermal refrigerator is thus smaller than the one characterizing a standard refrigerator operating between the warm and cold 
sources.
We choose for the present device $\mu_R=\mu_L=0$.

As detailed by Benenti {\it et al.} in Ref. [\onlinecite{Benenti17}], in a sequential framework, the cooling process can be schematized by some sequence 
among the two-dot states.
Labelling the states by 0, $b$, $t$ and $2$, respectively, for empty, bottom-dot occupied, top-dot occupied, and doubly occupied states, the cooling 
process corresponds to the following sequence: 0 - $b$ - 2 - $t$ - 0; hence, the electron coming from the top reservoir to fill the doubly occupied state 
borrowing the energy $\epsilon_t +U$, will reenter the same reservoir with an energy reduced by $U$.
In the same time an electron crosses the bottom dot, from the left to the right as a consequence of the choice of the functions $\Gamma_L$ and 
$\Gamma_R$.
This picture has been used in the SQME approach to delineate the parameter region where expecting the cooling regime. [\onlinecite{Benenti17}]. 
In SQME such a cycle leads to a total reservoir entropy variation of 
$\Delta S_0 =-\epsilon_b/T_L - (\epsilon_t+U)/T_t+ (\epsilon_b+U)/T_R +\epsilon_t /T_t$. 
For the sequence 0 - $b$ - 2 - $t$ - 0 to spontaneously occur, one needs $\Delta S_0 >0$, which leads to 
$\epsilon_b > U \frac {T_L}{T_t} \frac{ (T_R-T_t)}{(T_L-T_R)} \ge 0$.
This inequality was also laid out in Ref. [\onlinecite{Erdman18}], to ensure positive heat current streaming from the cold reservoir. 
We stress that this inequality is only for indicative purpose in the present paper, and 
as will be shown numerically later, this criterion is not sufficient to guarantee the refrigerator regime beyond the SQME framework.

Another guide for seeking the cooling regime can be drawn from the examination of the probabilities of the four two-dot states, respectively, $p_0$, $p_b$, $p_t$ and $p_2$. They are tied by the normalization sum: $p_0 +p_b+p_t +p_2=1$, and related to the dot occupancies: the mean number of electrons on the dots are $\langle n_b\rangle = p_b +p_2$, and
$\langle n_t\rangle = p_t +p_2$. The double occupancy is equal to the probability  of the doubly occupied state: $\langle n_b n_t \rangle =p_2$.
For the sequence 0 - $b$ - 2 - $t$ - 0 to occur, none of the four probabilities should be too low, in other words on-dot charge fluctuations must be as 
important as possible.
The parameter regime allowing the cooling operation is thus subject to competing requests: as suggested by entropy consideration in SQME,
one must have at least positive $\epsilon_b$ and so $\epsilon_b+U$: however, this leads to a $b$ dot with a low 
occupancy that is detrimental to charge fluctuations [\onlinecite{notefluctuation}]. As a consequence the desired 
cooling regime is rather narrow, and characterized by low performances as shown in the next figures. 
To alleviate these adverse effects, the modeling adopted in Ref. [\onlinecite{Erdman18}], which is otherwise the same as in the present paper, makes a major
different hypothesis. 
Erdman {\it et al.} do not presume {\sl a priori} any relation between what they call 
$\Gamma^{out/in}_\alpha (0)$ and $\Gamma^{out/in}_\alpha (1)$[\onlinecite{noteGamma}]. They choose them by optimization, and
in the present notations this leads to equalities between $\Gamma_t(\epsilon_t) \bigl( 1-f_t(\epsilon_t)\bigr)$, 
$\Gamma_t(\epsilon_t+U) f_t (\epsilon_t+U)$, $\Gamma_L(\epsilon_b) \bigl( 1-f_L(\epsilon_b)\bigr)$, 
and $\Gamma_R(\epsilon_b+U) f_R(\epsilon_b+U)$.
Their choice is advantageous for the all-thermal regime. 
In our model, it would require a tricky monitoring of the different dot-lead hybridization functions to achieve the preceding equality.
For the parameters explored in the present paper, with the choice of Eq.~(\ref{nosGamma}), we have up to three orders of magnitude between the preceding four tunneling terms.

\begin{figure*}
\begin{minipage}{0.49\linewidth}
\centerline{\includegraphics[width=1.0\linewidth]{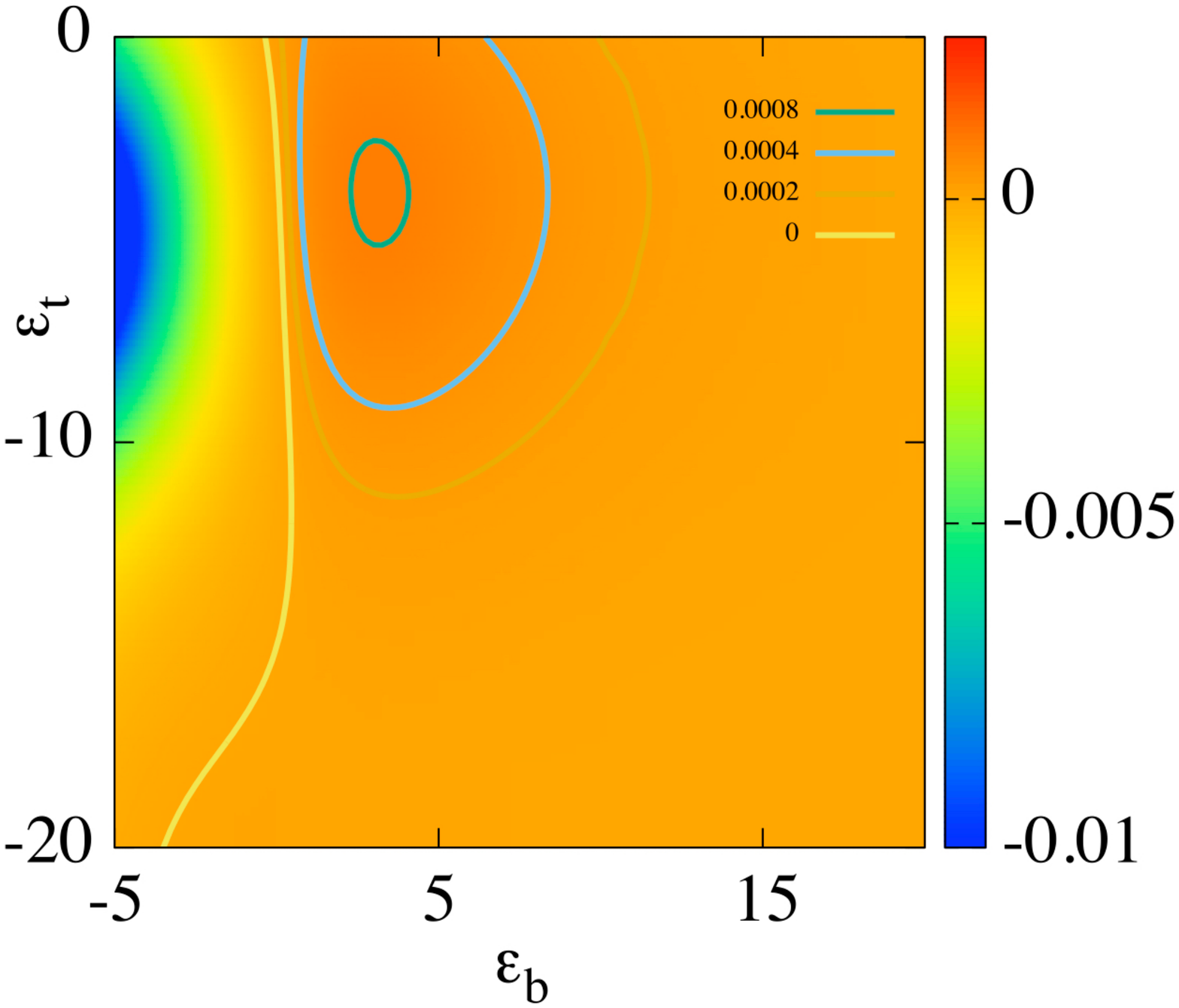}}
\end{minipage}
\hfill
\begin{minipage}{0.49\linewidth}
\centerline{\includegraphics[width=1.0\linewidth]{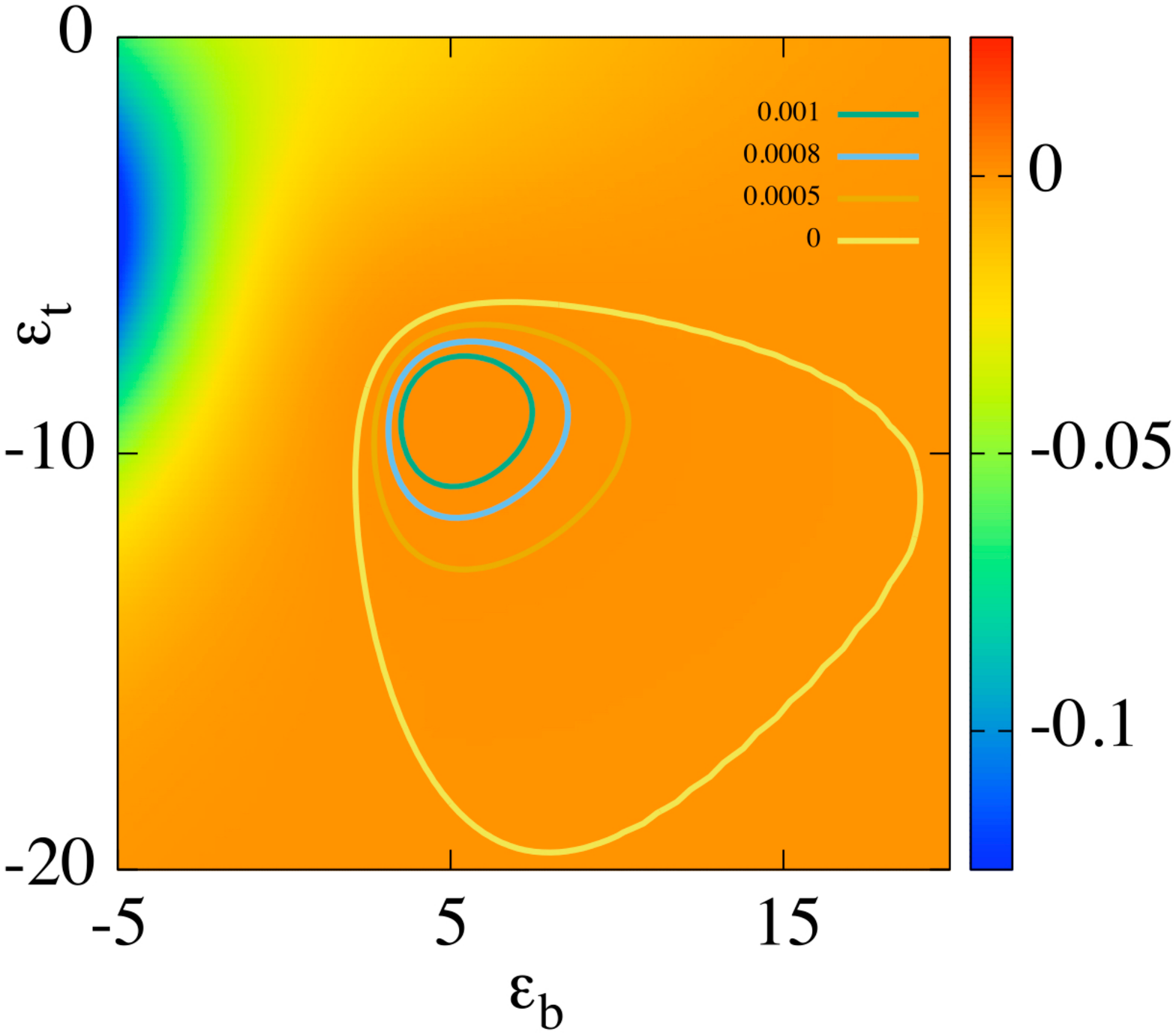}}
\end{minipage}
\caption{Maps of $J^e_b$ (left) and $J^Q_t$ (right) as functions of $\epsilon_b$ and $\epsilon_t$, for $U=10$, $T_L=4$, $T_R=2.1$, and $T_t=2$ in $\Gamma_b$ units. $J^e_b$ is in $e \Gamma/\hbar$ unit, $J^Q_t$ in $\Gamma^2/\hbar$ unit. Contour show the location corresponding to the cancelation of the currents. Other contours indicate also higher current values, respectively, ($2 \times 10^{-4}$, $4 \times 10^{-4}$, $8 \times 10^{-4}$) for charge and ($5 \times 10^{-4}$, $8 \times 10^{-4}$, 10$^{-3}$) for heat.}
\label{res-frig-all-ther}
\end{figure*}
In Fig.~\ref{res-frig-all-ther} the charge and heat currents of interest are plotted as functions of the $b$- and $t$-dot energies. A contour delimits the 
respective positive and negative regions, some current levels are also indicated. 
For the present parameters, the SQME entropy criteria would predict a cooling regime for $\epsilon_b$ exceeding 1.05. 
Furthermore in SQME, cooling power and electric current are proportionate and their ratio attains $U/e$. 
The left panel of Fig.~\ref{res-frig-all-ther} reveals
that the span of the cooling regime is much narrower, and also depends on $\epsilon_t$. Furthermore the signs of $J^Q_t$ and $J^e_b$ are not simply connected.
Finally, the ratio of $J^Q_t$ to $J^e_b$ varies and barely
reaches $ 5 \frac \Gamma e$ for the present parameters, in contrast to $U/e =10 \frac \Gamma e$ expected in SQME.
For the present parameters, we find a maximum electric current of $8.36 \times 10^{-4} \ \frac{e \Gamma}{\hbar}$, reached for $\epsilon_t =-3.8$, 
and $\epsilon_b=3$, whereas the maximum cooling power is $0.13 \times 10^{-2} \ \frac{\Gamma^2}{\hbar}$, for $\epsilon_t=-9$ and $\epsilon_b =5$. 
 A benchmark of the cooling power is the quantum bound per channel [\onlinecite{Whitney14, Whitney15, Whitney16}], which attains for the current
 parameters 0.52 $\frac{\Gamma^2}{\hbar}$, indicating that the maximum cooling power stays 
more than 400 times smaller than this value. 
The quantum bound $\frac{\pi^2}{12 h} k_B^2 T_t^2$, was found to be the maximum cooling power that can be extracted per channel through a device 
that can be described by a Landauer-type formula. It was shown that under some widespread hypothesis (non interacting leads and proportionate left and right 
lead-dot couplings) the Landauer current expression holds even for Coulomb coupled carriers [\onlinecite{Meir92}]. 
As a consequence, under the preceding suppositions, this bound is valid for one-dot two-terminal setup with Coulomb repulsion. However, in the present two-dot three-terminal geometry for which the assumptions are not fulfilled, the bound is only an indicative benchmark. 
Recently Luo {\it et al.} [\onlinecite{Luo18}]
established a bound for cooling power for interacting classical systems that is higher than the aforementioned one by a factor of $12/\pi^2$.
\begin{figure}
\begin{minipage}{0.49\linewidth}
\centerline{\includegraphics[width=1.0\linewidth]{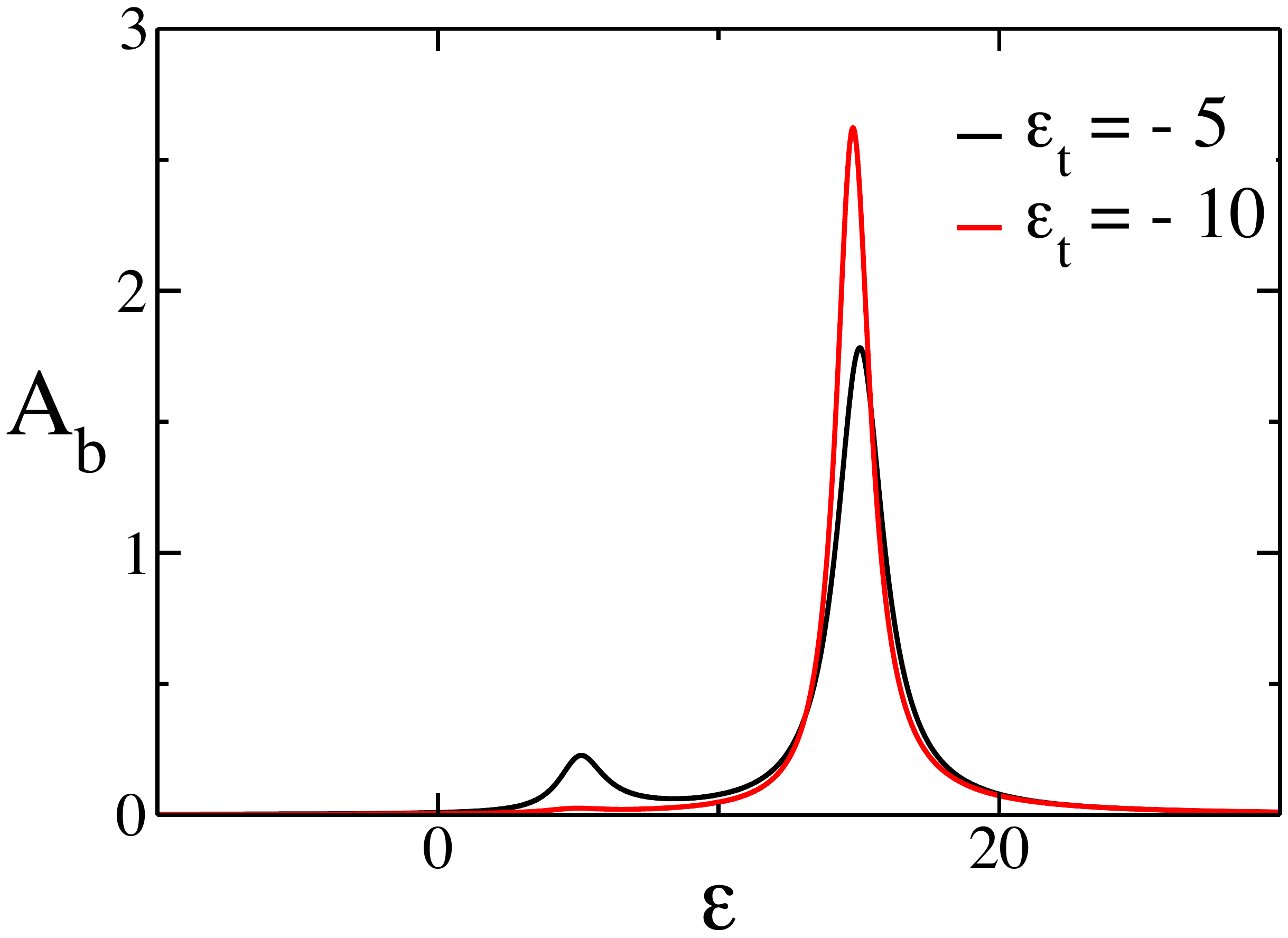}}
\end{minipage}
\hfill
\begin{minipage}{0.49\linewidth}
\centerline{\includegraphics[width=1.0\linewidth]{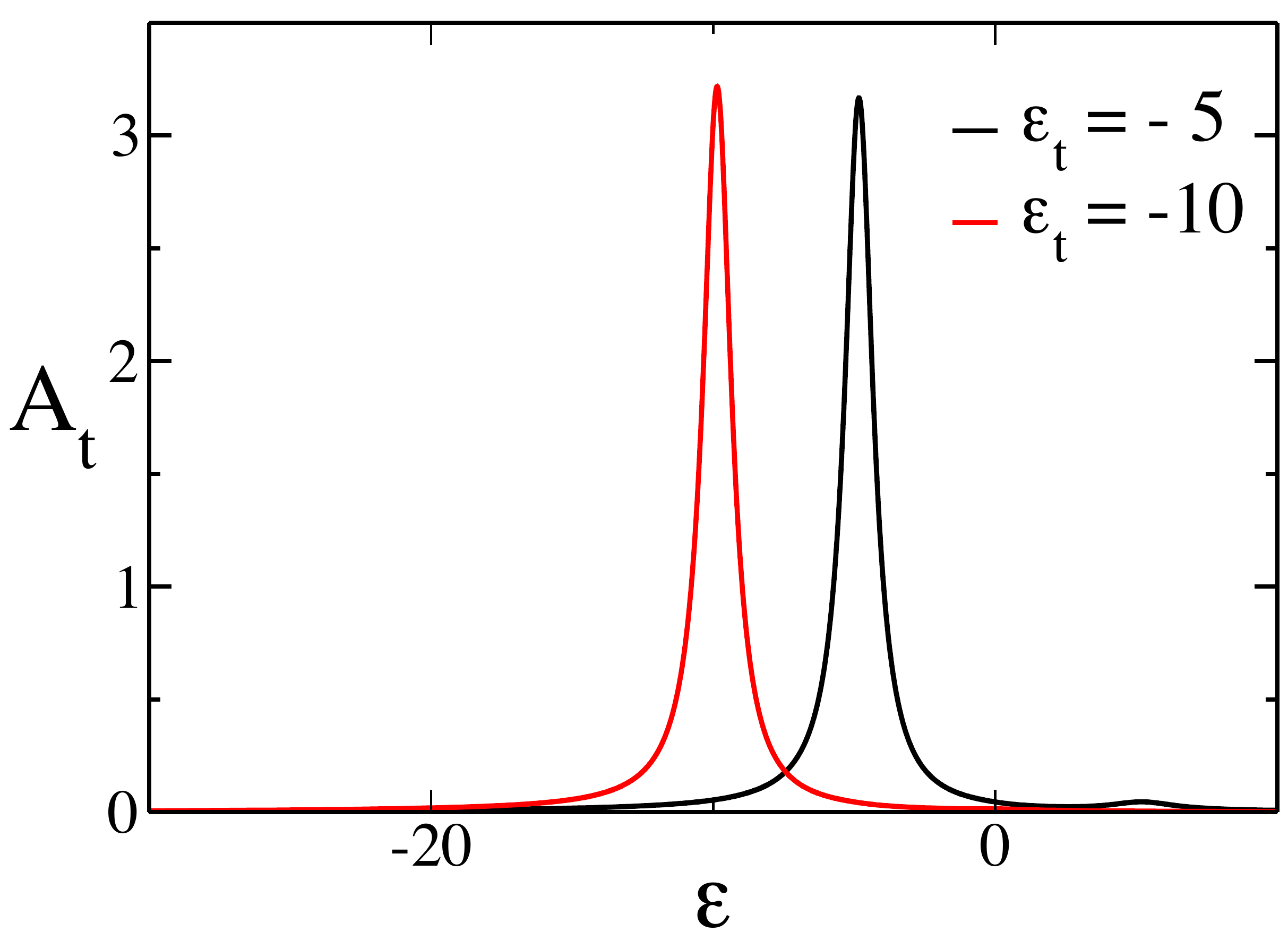}}
\end{minipage}
\caption{Dot spectral functions, for the same parameters as in Fig.~\ref{res-frig-all-ther}, for $\epsilon_b =5$ and two values of $\epsilon_t$. Left: spectral function of bottom dot. Right:  spectral function of top dot.}
\label{spectralfunctions}
\end{figure}

\begin{figure*}
\begin{minipage}{0.49\linewidth}
\centerline{\includegraphics[width=1.0\linewidth]{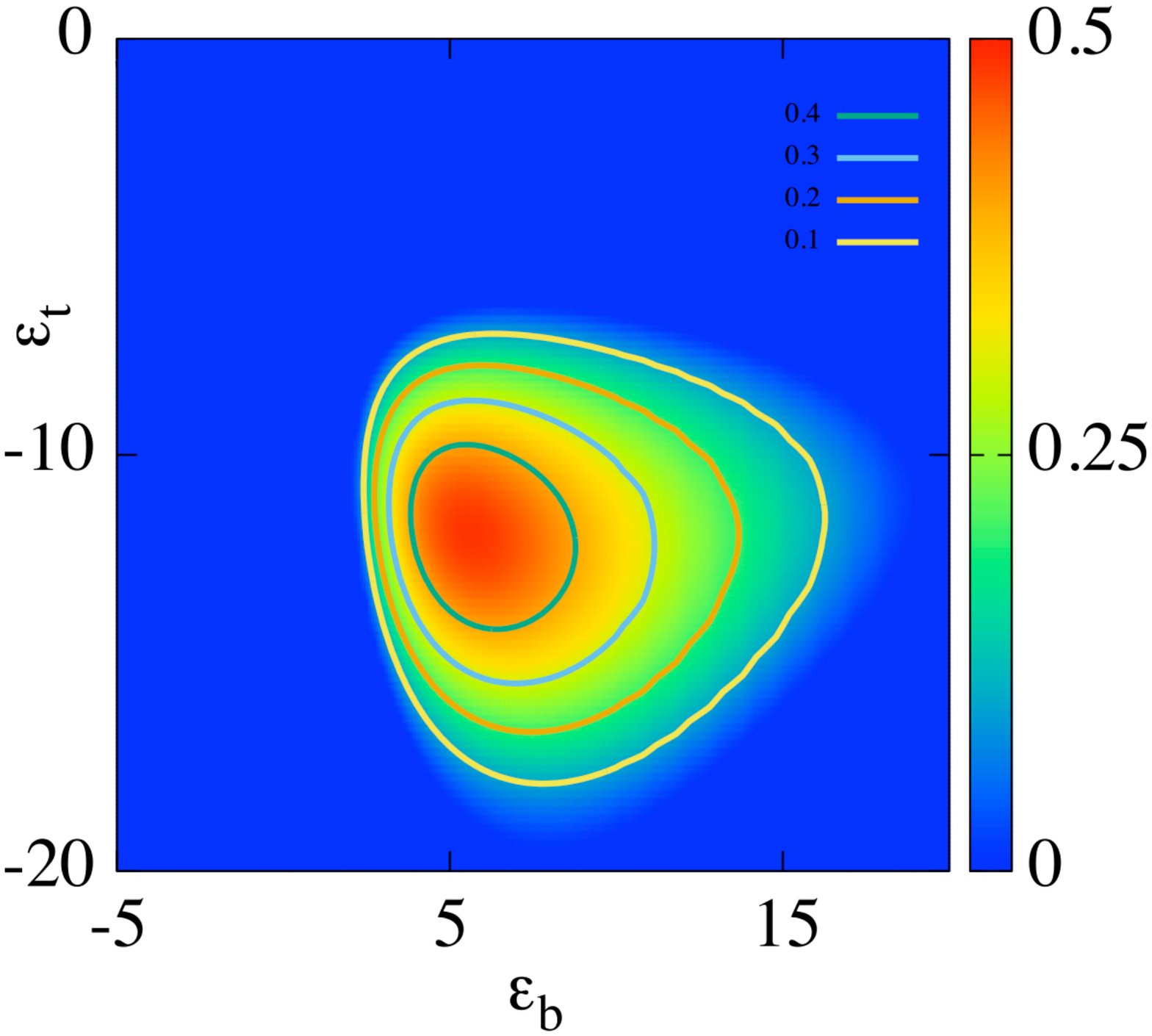}}
\end{minipage}
\hfill
\begin{minipage}{0.45\linewidth}
\centerline{\includegraphics[width=1.0\linewidth]{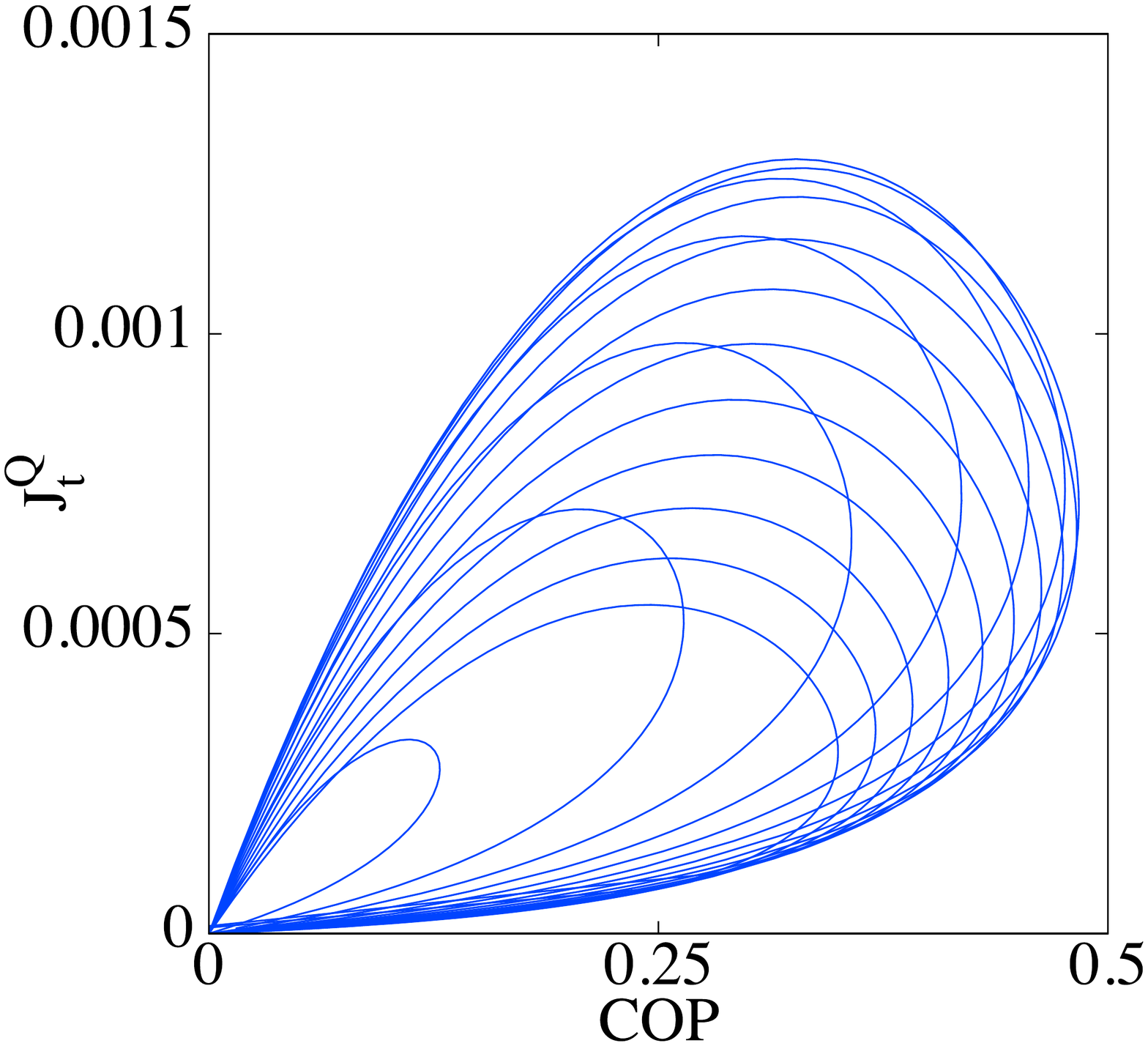}}
\end{minipage}
\caption{Left: map of the COP as function of $\epsilon_b$ and $\epsilon_t$ and some contours (0.1, 0.2, 0.3, 0.4). Right: $J^Q_t$ as a function of the COP for different values of 
$\epsilon_b$. The parameters are the same as in Fig.~\ref{res-frig-all-ther}.}
\label{COP}
\end{figure*}

It is not easy to predict the signs of heat and charge currents, except if the tunneling boundary $\epsilon_\Gamma=\epsilon_b +\frac U 2$ satisfies $\epsilon_\Gamma  < \mu_L $,  ($\epsilon_b<-5$ for the parameters of Fig.~\ref{res-frig-all-ther}). 
Then a positive $J^e_b$ would heat the $L$ reservoir, because the escape of an electron would correspond inevitably to a depletion under the chemical potential $\mu_L$.
However, this is forbidden by thermodynamics: $T_L$ being the highest temperature, in absence of any injected power, the $L$ source must cool down.
Thus for  $\epsilon_\Gamma  < \mu_L$, $J^e_b$ must be negative. 
To extend the discussion about charge and heat current signs for other parameter values, one has to resort to the approximation of narrow Green's function peaks.
Within this approximation one can discuss the cancelation of $J^e_b$ close to the line $\epsilon_b =0$.
If $\epsilon_b <\mu_L=0$ and $\epsilon_b + U > \mu_R=0$, $J^Q_L >0$ entails $J^e_b <0$: adding an electron under the chemical potential cools the $L$ lead, 
whereas the exit of the charge from the $R$ reservoir above its chemical potential cools it also. Thus $J^Q_R >0$, it follows that $J^Q_t<0$. 
For $\epsilon_b >\mu_L$ and still narrow Green's function peaks, $J^Q_L >0$ results in $J^e_b >0$ and $J^Q_R<0$. The preceding signs alone do not enable to fix the sign
of $J^Q_t$. 
The above discussion shows that $J^e_b <0$ triggers $J^Q_t <0$; we observed this result even when taking into account 
Green's functions with their finite width. However, positive $J^e_b$ does not bring $J^Q_t$ positivity.
With narrow peaks, $J^e_b$ cancels for $\epsilon_b =0$. From the left panel of Fig.~\ref{res-frig-all-ther}, the frontier appears a little bit displaced  due 
to the finite width of the Green's functions $G_b^\lessgtr(\epsilon)$, and influenced by the $\epsilon_t$ value.

The influence of $\epsilon_t$  onto the $b$-dot Green's functions can be noticed in Fig.~\ref{spectralfunctions} where the spectral function $A_b = i(G^>_b - G^<_b)$ is displayed as a function of energy for $\epsilon_b=5$, for the same parameters as in Fig.~\ref{res-frig-all-ther} and two different values of $\epsilon_t=-10,-5$. The peak positions, roughly located at $\epsilon_b$ and $\epsilon_b+U$, are slightly shifted by high $|\epsilon_t|$ value; more importantly the peaks amplitudes are modified. 
For the present parameters, one has  $A_b$ quite similar to $i G^>_b$, this is related to the low $b$-dot occupancy.
The corresponding $t$-dot spectral function is also presented in the right part of Fig.~\ref{spectralfunctions}. Due to the high $t$-dot occupancy, one has $A_t$ fairly close to $- i G^<_t$.
Finally, when $\epsilon_b$ raises, $J^e_b$ eventually decreases; it is simply related to the Fermi function behavior: as $\epsilon_b$ rises, less charges
are available in the $L$ reservoir to flow through the device.

In Fig.~\ref{COP}, a map of the COP for the same parameters as in Fig.~\ref{res-frig-all-ther} is displayed (left), together with 
a graph of the cooling power as a function of the COP (right).
The maximum COP is achieved for $\epsilon_t=-12$, and $\epsilon_b=5.75$, and attains 0.484, barely more than $1/20$ of the 
reversible one equal to 9.5.
In the right panel of the figure, the different curves correspond to different $\epsilon_b$ values, and are roamed clockwise as $\epsilon_t$ decreases.
Following one curve, it can be seen that the maximum cooling power and the maximum COP do not coincide, requiring a compromise in operating this 
kind of refrigerator. 
Adopting a tunnel coupling value of $\Gamma \simeq 20 \ \mu$eV,  compatible with the experimental value reported in Ref. [\onlinecite{Thierschmann15-1}], 
the present parameters correspond to $T_L \simeq 1$ K, and $U=0.2$ meV. The maximum charge current reaches about 4 pA, 
and the maximum cooling power hits 0.13 fW.
At the maximum cooling power, the COP is 0.328, and we find the following probabilities describing the two-dot states: $p_b =4 \ 10^{-3} ,\  p_t=0.952 , \ p_2=0.013,$ and
$p_0=0.031$. The low $p_b$ value is probably connected to the low performances.

The all-thermal refrigerator regime is suppressed if source temperatures are significantly reduced. 
We interpret it as a lack of on-dot charge fluctuations that get even smaller than the previous ones when temperatures lower. 
The energy current values in the all-thermal device are very sensitive to the difference $(T_R-T_t)$. The best performances are obtained for $T_R=T_t$, for which, 
for the same parameters as previously, except $T_R=2$, the 
maximum cooling power nearly doubles compared to the previous case, and the COP attains 0.65.
Meanwhile, for $T_R-T_t=0.4$, $T_t=2$, and $T_L=4$, the cooling regime is very narrow in the ($\epsilon_b, \epsilon_t$) space and underperforming: 
achieving a maximal COP around 0.1, and a maximum cooling power close to 10$^{-4}$ in $\Gamma^2/\hbar$ units. 
The influence of $(T_R-T_t)$, keeping $(T_L-T_t)$ unchanged, can be understood from entropy consideration.
For the all-thermal refrigerator, the last term of Eq. (\ref{dotS0}) cancels, and it can be seen that in the cooling regime ($J^Q_t >0$), 
$T_R \rightarrow T_t$ has a positive effect on the two terms of the entropy production rate: enhancing the positive contribution and reducing the negative one. 
Experimentally it can be advantageous: thermal insulation can be tricky at the nanoscale, but it appears that a bad thermal insulation between $R$ and $t$ 
reservoirs can be favorable. 
The case $T_R=T_t$ may sound paradoxical: in this case the all-thermal refrigerator is only a two-temperature machine without any injected power.
Extracting heat from the cold $t$ reservoir is nevertheless possible and may seem to be violating the second law statement.
Obviously the paradox is solved by accounting for the whole cold bath made of the $R$ and $t$ reservoirs, that globally gains heat from the $L$ hot source.  

\section{Electric refrigerator}

\begin{figure}
\begin{minipage}{0.49\linewidth}
\centerline{\includegraphics[width=1.0\linewidth]{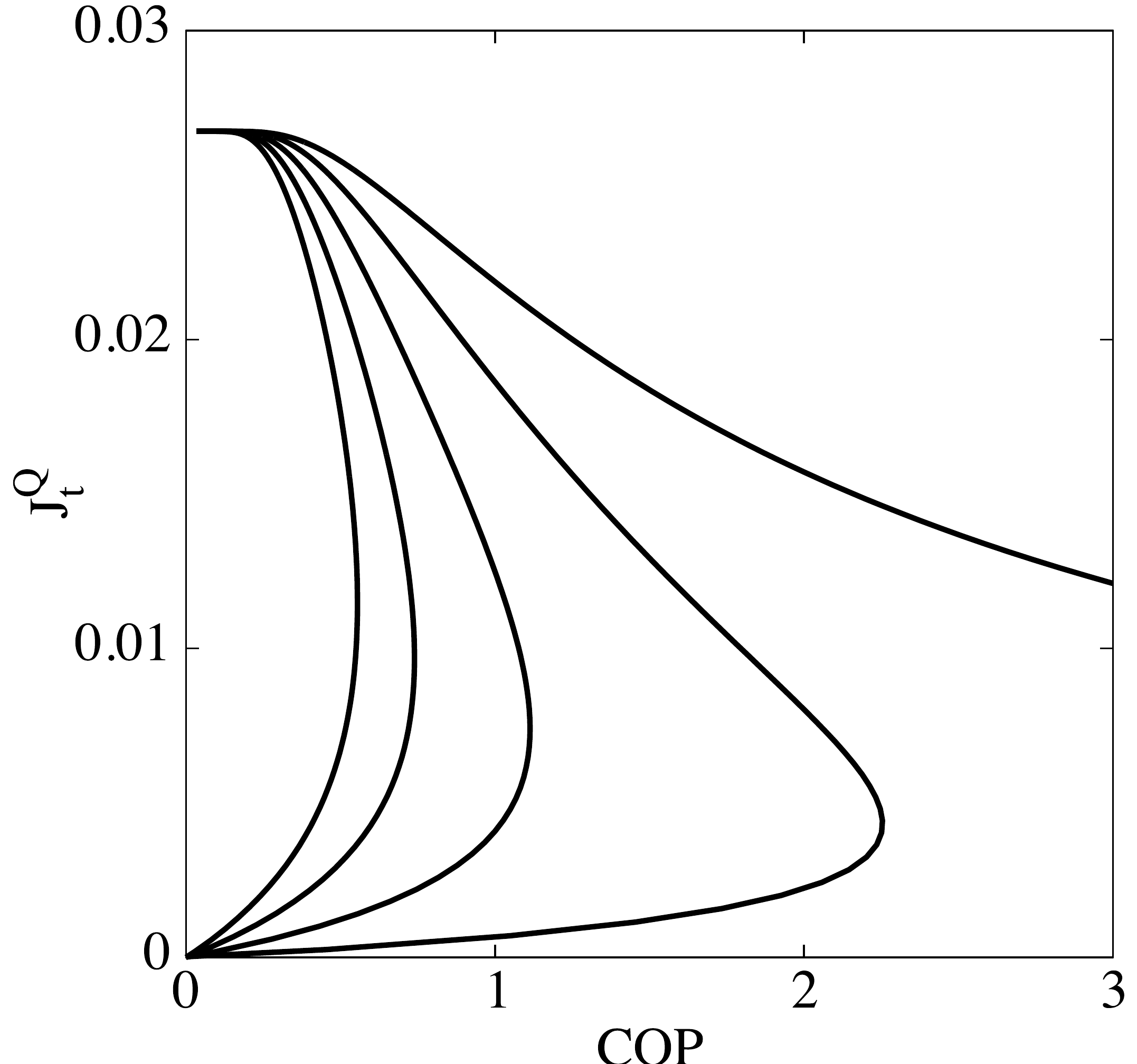}}
\end{minipage}
\hfill
\begin{minipage}{0.49\linewidth}
\centerline{\includegraphics[width=1.0\linewidth]{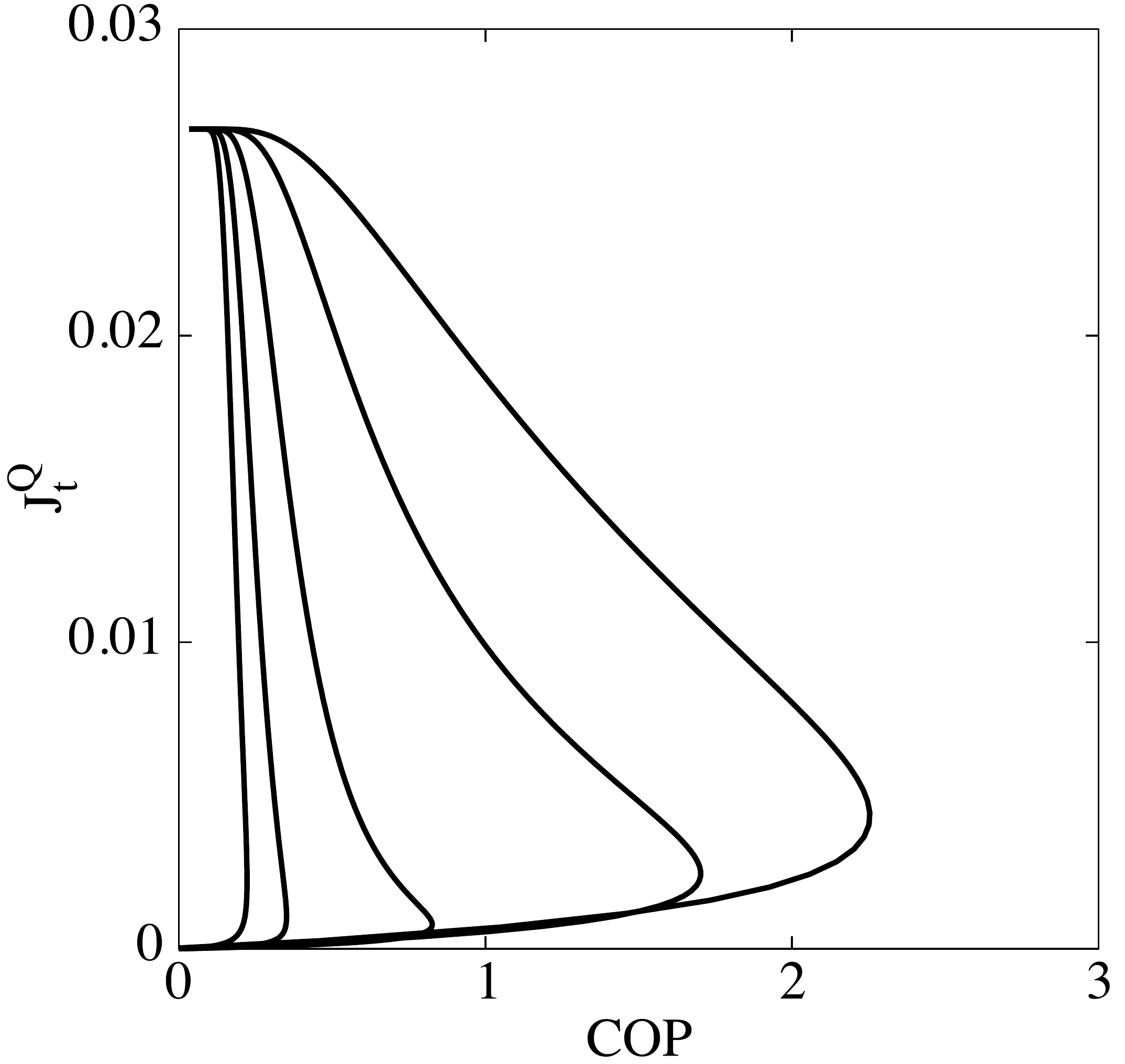}}
\end{minipage}
\caption{ Cooling power in $\Gamma^2/\hbar$ unit, as function of the COP, for $U=10$, $\epsilon_t=-5$, $T_t=1$, $T_b =T_t +\Delta T$. Left:
for different values of $\Delta T =0, 0.1, 0.2, 0.3, 0.4$, and 
$\epsilon_b=-5$. Right: for different values of $\epsilon_b=-5, -2.5, 0, 2.5, 5$, and $\Delta T=0.1$. See text for further explanation.}
\label{frigelec1}
\end{figure}

We turn to the case where cooling of the cold $t$ reservoir is monitored by a voltage bias $V$ applied between the two bottom sources. We apply it symmetrically, 
choosing $\mu_L = -\mu_R =  \frac{V}{2}$, and adopt the following notations:
$T_L=T_R =T_b=T_t+\Delta T$. The COP is defined in the present case by the ratio: $J^Q_t/(J^e_b \times V)$ and bounded by the reversible one ${T_t}/{(T_b-T_t)}$.
Our calculations establish that choosing $\epsilon_t =\epsilon_b=- \frac U 2$, which corresponds to half-filled dots, is advantageous for the charge current and cooling power: all other factors being equal,
charge and thermal currents as well as COP are higher, due to the favorable on dot fluctuations at half-filling [\onlinecite{notefluctuation}].
In Fig.~\ref{frigelec1} the cooling power is displayed as a function of the COP. The different curves correspond to different values of $\Delta T$ (left) or to different 
values of $\epsilon_b$ (right). Along all the different lines, $J^Q_t$ and $V$ raise concurrently. 
They were obtained for $U=10$, $\epsilon_t=-5$, and $T_t=1$.
For the left plot one has $\epsilon_b=\epsilon_t$, and for the same value of $J^Q_t$, raising $\Delta T$ lowers the COP.
This behavior can be enlightened by the following remarks concerning $J^Q_t(V)$ and $J^e_b(V)$: as will be discussed soon, for $V=0$,
$J^Q_t$, and $J^e_b$ are negative, and a higher $\Delta T$ leads to a higher $|J^Q_t|$ as expected for heat transfers between different temperature sources. This is 
 achieved by a $J^e_b$ that grows also with $\Delta T$ in absolute value. 
 $J^Q_t$  and $J^e_b$ increase with $V$, and a finite
voltage bias eventually reverses the signs of $J^e_b$ and $J^Q_t$ such that the machine switches to the refrigerator regime. 
However, $J^Q_t$  and $J^e_b$ stay lower for higher $\Delta T$ because of their lower $V=0$ starting point. 
For $\Delta T=0$, the COP is infinite at $V=0$ because charge and heat currents cancel proportionately. 
For the right plot of Fig.~\ref{frigelec1}, $\Delta T =0.1$, and for the same $y$ value, 
the COP gets lower as $\epsilon_b$ moves away from $-\frac U 2$.
The reversible COP for the parameters of the right part of Fig.~\ref{frigelec1} is 10.
 
\begin{figure}
\centerline{\includegraphics[width=0.7\linewidth]{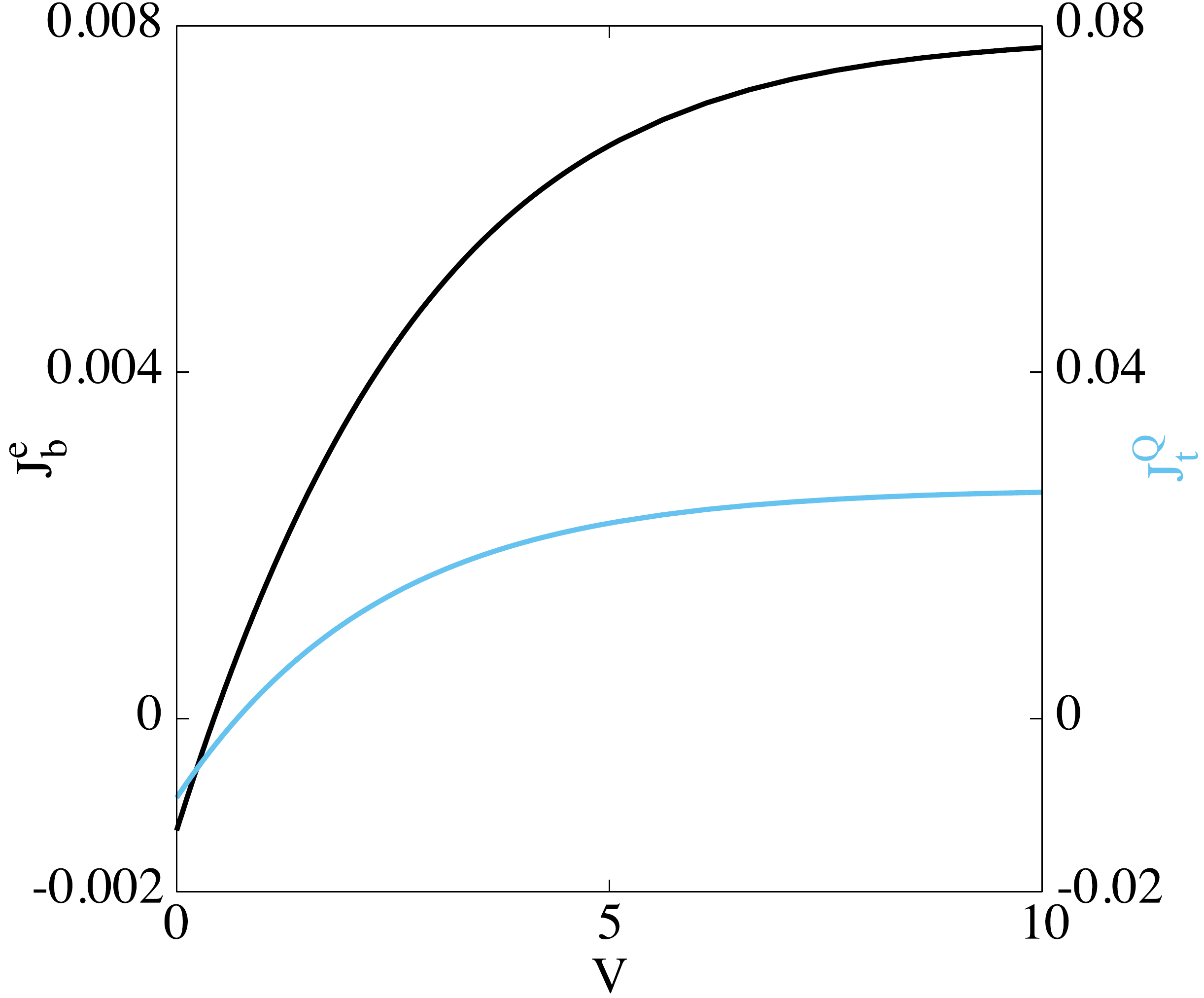}}
\caption{ Electric current (black, left scale) and cooling power (blue, right scale) as a function of $V$, for $U=10$, $\epsilon_t=\epsilon_b =-5$, $T_t=1$, and $T_b=1.1$.}
\label{frigelec3}
\end{figure}

In both panels of Fig.~\ref{frigelec1} it appears that $J^Q_t$ saturates at high bias, the same behavior is observed for $J^e_b$ as can be seen in Fig.~\ref{frigelec3}. 
This is easy to unravel in the case of $J^e_b$:
the $L$-Fermi function differs from its zero-temperature values by less that 2\% when moving away from the chemical potential $\mu_L=\frac {eV}{2}$, by 
$4 T_b$ on both sides.
Then for $ \epsilon_\Gamma \lessapprox  \mu_L -4 T_b  $, that is with the present parameters $ V \gtrapprox 8 T_b = 8.8$, the charge current flowing in between the $L$ lead and the $b$ dot will not really depend 
anymore on the bias. 
The same argument applies to $J^Q_R$ and $J^Q_L$, and as a consequence to $J^Q_t$.
The cooling power and electric current saturation values do not depend on $\Delta T$, nor on 
$\epsilon_b$: they only depend on $U$ and $T_t$ (we choose $\epsilon_t=-U/2$). 
In this last figure, it is obvious that heat and electric currents are not proportionate: the ratio is lower than predicted by SQME as previously discussed, due to cotunneling and 
higher-order processes. 

In Fig.~\ref{frigelec3} we observe that both currents are negative at null bias. For $J^Q_t$ it is a consequence of the second law. Indeed in absence of any input power, the cold source can only get warmer, leading in our convention to $J^Q_t <0$. 
Then as the voltage bias raises, the current signs will eventually change. We can understand that the first quantity to cancel is the electric current.
Indeed, from the Eq. (\ref{dotS0}), adjusted to the present case and notations, one has
\begin{equation}
\dot{S}_0 = J^Q_t \Bigl(  \frac{1}{T_b}-\frac{1}{T_t}\Bigr) +\frac {V J^e_b}{T_b} \ .
\end{equation}
For $V >0$, as long as $J^e_b \le 0$, $J^Q_t$ must be also negative such as to guarantee the positivity of the entropy production rate.
A situation with a positive $J^e_b$ and a negative $J^Q_t$, as was already observed in some parameter range of the all-thermal refrigerator, relies on the breadth of the Green's functions. 
At the specific voltage for which $J^e_b=0$, charge current fluctuations reduce the heat flow between hot and cold sources by a factor around 3 compared to the null bias situation.
When the bias is such that $J^Q_t=0$ whereas $J^e_b > 0$, the two dots not only do not share any charge, but also no energy on average.
However, current fluctuations as earlier are at work, such that the dots are not independent from each other as witnessed by the finite current $J^e_b$.
In brief, with broadened Green's functions, cooling power and electric current probably do not cancel for the same bias. 
However, thermodynamics prevents $J^Q_t$ to cancel as long as $J^e_b$ is negative.

For $U=10$, $\epsilon_b =\epsilon_t =- \frac U 2$, $T_t=1$, and $\Delta T =0.1$, 
the asymptotic charge current reaches $7.92 \times 10^{-3} \frac {e \Gamma}{\hbar}$, whereas the asymptotic heat current is 
$2.67 \times 10^{-2} \frac { \Gamma^2}{\hbar}$. This last value can be compared to the bound predicted by Whitney [\onlinecite{Whitney14, Whitney15}] for a one-channel two-terminal setup which attains 0.13 in the same units, making the electric refrigerator significantly higher performing than the all-thermal one, with a ratio $J^Q_{asym}/J_{qb}^Q$ close to 1/5 
($J^Q_{max}/J_{qb}^Q$ was close to 1/400 for the all-thermal machine).
\begin{figure}
\begin{minipage}{0.49\linewidth}
\centerline{\includegraphics[width=1.0\linewidth]{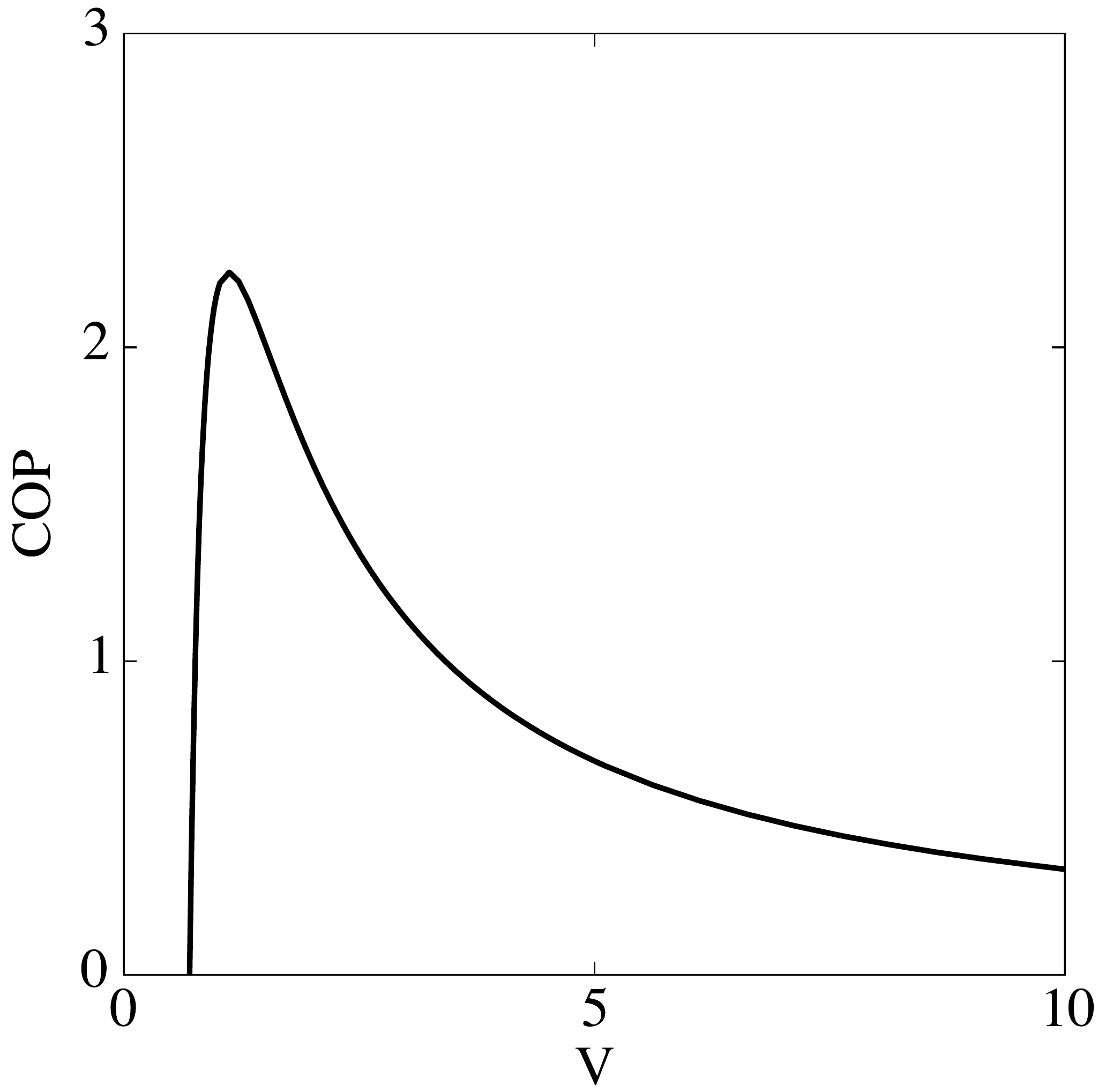}}
\end{minipage}
\hfill
\begin{minipage}{0.49\linewidth}
\centerline{\includegraphics[width=1.0\linewidth]{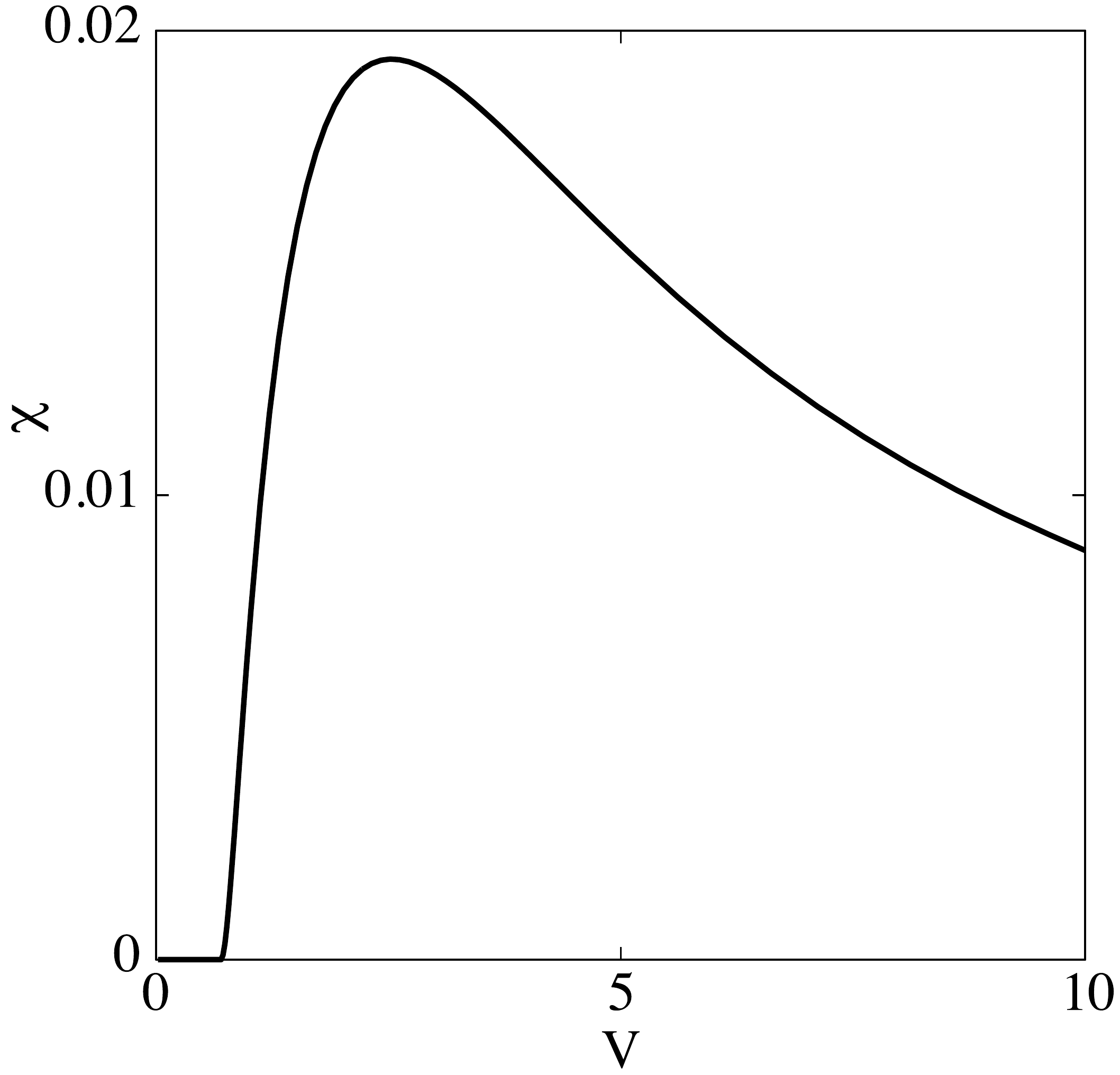}}
\end{minipage}
\caption{ COP (left) and $\chi$ (right) as a function of $V$ for $U=10$, $\epsilon_t=\epsilon_b =- U/2$, $T_t=1$, $T_b=1.1$.}
\label{frigelec5}
\end{figure}
However, the asymptotic cooling power is not very interesting, due to the corresponding null COP. 
The Fig.~\ref{frigelec5} (left) completes the picture by showing a graph of the COP as a function of $V$ for the same parameters as in Fig.~\ref{frigelec3}. 
The COP increases in an abrupt way before attaining its maximum at low bias, reaching 2.24 (the reversible one is 10), and afterwards scales as $1/V$ at high bias as a consequence of both current 
saturations. 
In contrast to the engine, for which maximum output power and maximum efficiency nearly coincide (see Fig. 4 of Ref.~[\onlinecite{Dare17}]), 
a compromise has to be found in selecting an operating point for the present thermal machine:
at the maximum COP, the cooling power is only 17\% of its asymptotic value.
A way to select the operating point of this refrigerator, is to use the $\chi$ criterion [\onlinecite{Yan90,Tomas12}]. The $\chi$ function is defined as 
$\chi = \mathrm {COP} \times J^Q_t$, and is plotted in Fig.~\ref{frigelec5} (right) as a function of bias for the same parameters as in the left panel. 
At maximum $\chi$, the COP still reaches 1.33, whereas the cooling power is equal to 55\% of its asymptotic value, the charge current attains also 55\% of its asymptotic value.
With $\Gamma \simeq 20 \ \mu$eV, the currents flowing through the device at maximum $\chi$ are  the following: a charge current of 22 pA, and a cooling power of 1.4 fW.

For the parameters of Fig.~\ref{frigelec3}, the two dots are half-filled as a consequence of $\epsilon_b =\epsilon_t =-U/2$. 
As $V$ raises, the probabilities of the different two-dot states are stable from low to high bias: $p_t=p_b \simeq 0.46$,
whereas $p_2=p_0 \simeq 0.04$: the present situation is different from the one encountered for the all-thermal 
refrigerator machine, where $p_t$ was much lower. 

\section{Conclusions}

Using a formalism that was set up for strongly correlated systems [\onlinecite{U-comment}], which fulfills the first and second principles, we have presented a 
comparative study between two types of refrigerators, 
one of them being powered by heat, the other one by electric supply.
The latter is rather competitive in terms of cooling power, which reaches a significant fraction of the quantum bound.
This study shows that for the same reservoir properties, the all-thermal refrigerator is much less competitive and is limited in its operating regime.
The reason for these underperforming properties resides in the lack of on-dot charge fluctuations.
However these might be probably magnified by reservoir engineering as proposed in Ref.~[\onlinecite{Correa14,Correa14-2}]. 
Modifying bath properties by DOS or hybridization tailoring deserves to be explored and may be compatible with the NCA technique. This is left for further studies.

The case of the power driven refrigerator confirms the primacy of the three-terminal geometry over the two-terminal one. Indeed attaining an appreciable fraction 
of the cooling power quantum bound per channel, as obtained in this paper, is not guaranteed. This was also pointed out for the three-terminal two-dot engine 
where the output power achieved a substantial part of the corresponding quantum bound [\onlinecite{Dare17}]. 
The present setup takes benefit from the three-terminal geometry [\onlinecite{Mazza14,Sartipi18}] combined with a favorable effect of 
electronic correlations [\onlinecite{Luo18}].

\section*{Acknowledgment}
The author gratefully acknowledges P. Lombardo, R. Marhic, and A. Deville for valuable discussions and suggestions.


\begin{thebibliography}{99}
\bibitem{ambiguite} Designating an  all-thermal machine as autonomous one may be ambiguous, this term being also used sometimes for indicating 
machines operated in the steady-state regime [\onlinecite{Mitchison19}].
\bibitem{notefluctuation} The $t$-dot occupancy fluctuations are maximum at half-filling: 
$\langle (n_t -\langle n_t \rangle)^2\rangle = \langle n_t \rangle (1-\langle n_t \rangle)= (p_t+p_2)(p_b+p_0)$. There is a symmetric expression for the 
$b$-dot occupancy.
\bibitem{noteGamma} The correspondence is the following $\Gamma_t^{out}(0)= \Gamma_t(\epsilon_t) \bigl( 1-f_t(\epsilon_t)\bigr)$, 
$\Gamma_t^{in}(1)=\Gamma_t(\epsilon_t+U) f_t (\epsilon_t+U)$,
$ \Gamma_L^{out} (0)= \Gamma_L(\epsilon_b) \bigl( 1-f_L(\epsilon_b)\bigr)$, and
$ \Gamma_R^{in} (1)= \Gamma_R(\epsilon_b+U) f_R(\epsilon_b+U)$.
\bibitem{U-comment} Small $U/\Gamma$ values are not explored in the present calculations, the NCA being not appropriate when this ratio is low.
Though, higher $U/\Gamma$ values would be detrimental to transport properties due to an adverse effect on charge fluctuations.
\end{thebibliography}
\end{document}